\documentclass{iaus}
\usepackage{graphicx}

\title[Nebulae around WR stars] 
{High-excitation nebul\ae\ around Magellanic Wolf-Rayet stars}

\author[Pakull]   
{Manfred W. Pakull$^1$
}

\affiliation{$^1$CNRS, Observatoire Astronomique,
11 rue de l'Universit\'e, F67000 Strasbourg, France\\
email: {\tt pakull@astro.u-strasbg.fr} \\
}

\pubyear{2008}
\volume{256}  
\pagerange{}
\jname{The Magellanic System: Stars, Gas, and Galaxies}
\editors{Jacco Th. van Loon \& Joana M. Oliveira, eds.}
\begin{document}

\maketitle

\begin{abstract}
The SMC harbours a class of hot nitrogen-sequence Wolf-Rayet stars (WNE)
that  display only relatively weak broad 
emission lines. This indicates low mass-loss rates and makes them  
also hard to detect. However, such stars are possible
emitters of strong He$^+$ Lyman continua that in turn could ionize
observable He\,{\sc iii} regions, i.e. highly excited H\,{\sc ii}
regions emitting nebular He\,{\sc ii}\,$\lambda$4686\ emission. We here
report the discovery of a rare He\,{\sc iii} region in the SMC which is
located in the OB association NGC\,249 around the weak-lined WNE star 
SMC\,WR10. WR10 is particularly interesting since it is a single star
showing the presence of atmospheric hydrogen. While analysing the
spectrum in the framework of two popular WR atmosphere
models, we found for the same input parameters strongly discrepant 
predictions (by 1 dex) for the He$^+$ Lyman continuum . A second 
aspect of the work reported here concerns the beautiful MCELS color
images which clearly reveal a class of strongly 
\mbox{[O\,{\sc iii}]\,$\lambda$5007} emitting (blue-coded) nebulae. Not unexpectedly,
most of the  'blue' nebulae are known Wolf-Rayet bubbles, but new
bubbles around a  few WRs are also detected. Moreover, we report the
existence of blue nebulae without associated known WRs and discuss the
possibility that they reveal weak-wind WR stars with very faint stellar
He\,{\sc ii}\,$\lambda$4686 emission. Alternatively, such nebulae might
hint at the hitherto missing population  of relatively low-mass hot He
star components predicted by massive binary evolution calculations. Such
a binary system is probably responsible for the ionization of the bright
He\,{\sc ii}\,$\lambda$4686-emitting nebula N\,44C.
\end{abstract}

\firstsection 

\section{Wolf-Rayet stars and their He$^+$ ionizing radiation}

My interest for high-excitation H\,{\sc ii} regions dates back to the 
late 1980s. At that time we tried to understand the formation of  the
extended He\,{\sc ii}\,$\lambda$4686 recombination emission around  the 
massive black-hole candidate LMC\,X-1 (\cite[Pakull \& Angebault
1986]{PA86})  in terms of photoionization and to draw conclusions about 
the otherwise unobservable EUV/soft X-ray ionizing continuum of the
X-ray source. X-ray ionization has been a  well-understood process, but
hitherto mainly applied to power-law ionizing continua in AGNs.  It
turned out that the only known examples of nebular  $\lambda$4686
emission in H\,{\sc ii} regions concerned two Wolf-Rayet (WR)  ring
nebulae, one around the presumably very hot Galactic WO star WR102 and
the other  one surrounding a similar object in the Local Group galaxy
IC\,1613  (see, e.g., \cite[Pakull \& Motch 1989b]{Pa89a} and references
therein). Another prominent He\,{\sc iii} region is also located in the
LMC, namely the small  region N44C (= NGC\,1936) within the larger
H\,{\sc ii} complex N44. Here the  ionizing star is not a WR, but
appears to be a normal O7 star. This led Pakull \& Motch (1989a)
\cite[]{PaMo89} to propose an  interpretation in terms of a fossil X-ray
ionized nebula, possibly due to  the transient source LMC\,X-5.   As a
historical remark, I'd like to mention that the presence of  highly
ionized He\,{\sc iii} regions around the two WO stars led
\cite[Terlevich \& Melnick (1985)]{TM85} to suggest that a population
of  very hot ($T_{\rm rad} >$ 80--100 kK) luminous stars might well be
responsible  for the high ionization observed in AGNs, thus challenging
the general  interpretation in terms of massive accreting black holes. 
For such stars they coined the fitting term {\it Warmers}.
 
In 1991, at the Bali IAU Symposium 143 on Wolf-Rayet stars two groups 
(\cite[Niemela et al. 1991]{Nie91}; \cite[Pakull 1991]{Pa91}; 
\cite[Pakull \& Bianchi 1991]{Pabi91}) independently announced the
discovery of He\,{\sc iii} regions around  a few Magellanic WR stars.
They were SMC\,WR7 (located in N\,76) in the SMC, and in the LMC Brey\,2
(located in N\,79) and Brey\,40a  (within an anonymous H\,{\sc ii}
region).  The spectral types of these  Magellanic 'Wolf-Rayet Warmers'
(hereafter WRW) came as a surprise:  these stars are early WN types
rather than the more advanced  WO types, as one might have naively
expected from the previously  recognized objects. Moreover, it was shown
that the two known Magellanic WOs do not excite observable He\,{\sc iii}
regions (\cite[Pakull 1991]{Pa91}), i.e.,  a WO is not necessarily a
WRW. Since then, and although several additional Magellanic WR stars
have been  discovered in the meantime, no new WRW turned up in the MCs.

As a result of the pioneering work by  \cite[Schmutz, Leitherer \&
Gruenwald (1992)]{SLG92}, we know that the ionizing radiation from WR
stars does not only depend on the atmospheric  'core' temperature of the
star (i.e., at a radius where the wind is still  subsonic), but also on
the mass-loss rate. In short, a sufficiently strong wind from a  WN star
(consisting mainly of helium) will absorb all He$^+$ Lyman continuum 
photons (i.e., $h\nu>54$ eV), not unlike an ionization-bound
Str\"{o}mgren sphere.  If however the wind is sufficiently weak, a large
fraction of the hard ionizing photons will escape and create observable
He\,{\sc iii} regions in the surrounding ISM. Grids of the most recent
Wolf-Rayet star models,  which among others also take into account
line-blanketing effects are described  in \cite[Smith et al.
(2002)]{Smetal02} (CMFGEN models) and in  \cite[Hamann \& Gr\"afener
(2004)]{HaGr04} (PoWR models).  Note that the CMFGEN grid is implemented
in the popular Starburst99  software package. As will be shown below,
and contrary to what is sometimes assumed, the two models strongly
disagree on the predicted power of the emitted  He$^+$ Lyman continuum
emission.    

The study of Wolf-Rayet stars currently witnesses a renaissance since it
was realised that these stars -- or rather a not-yet-specified rare
subclass thereof -- appear to be the direct progenitors of (long) gamma
ray  bursts  (cf., \cite[Yoon \& Langer 2005]{YL05}). One important
ingredient  of currently  favoured models is a very high rotation rate
of the core at the time of collapse/jet formation. However, such a
configuration is not easily realized due  to expected previous angular
momentum loss via the stellar wind.

At low metallicity, WR winds are now known to be weaker than those of
their more metal-rich  Galactic counterparts (cf., \cite[Vink \& de
Koter 2005]{ViKo05}  and references therein), and it is possible  that
under these circumstances  (low $Z$, very rapid rotation) massive
stellar evolution proceeds quasi-chemically homogeneously  (\cite[Yoon
\& Langer 2005]{YL05}). In this scenario, the stars remain in the blue
part of the Hertzsprung Russell (HR) diagram  and evolve from the main
sequence directly bluewards to high effective temperatures ($T_{\rm
eff}>$ 70--100 kK), even if their outer layers still contain hydrogen.
It is however an open question whether homogeneous evolution is indeed
realized in nature.

\section{SMC\,WR10: A second Warmer in the Small Cloud}

Shortly after the somewhat unexpected discovery  of several new  WR
stars in the SMC (\cite[Massey \& Duffy 2001]{MaDu01})  we observed
long-slit spectra of the newly identified WR10  which is located within
the H\,{\sc ii} region NGC\,249   (Fig. 1). Although this object has
subsequently also  been observed with various similar instrumental
set-ups, no other observer seems to have noticed (or have cared to look
at) the relatively strong {\it extended} He\,{\sc ii}\,$\lambda$4686
emission that is depicted in Fig. 2.

\begin{figure}[t!]
\begin{center}
 \includegraphics[width=4.in]{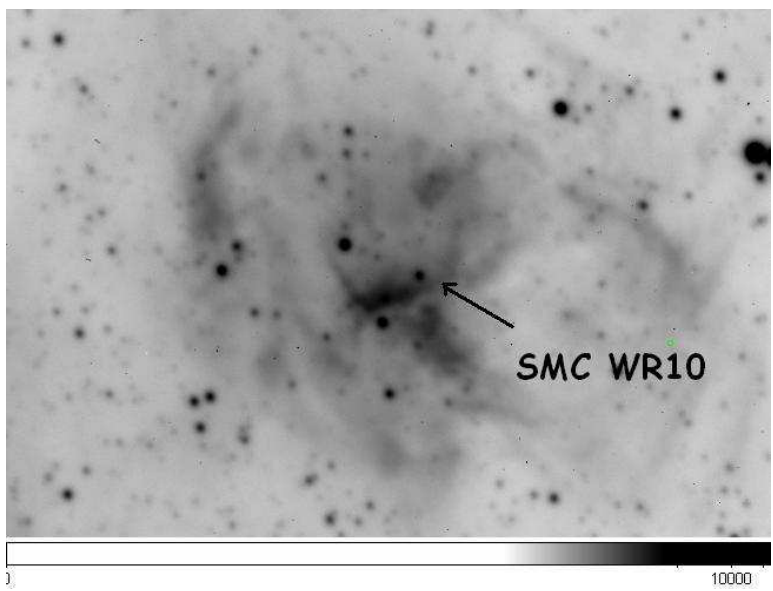} 
 \caption{\textit{ESO} \textit{NTT}/EMMI H$\alpha$ image of the 
 SMC H\,{\sc ii} region NGC\,249 in which the 
 Wolf-Rayet star SMC\,WR10 is embedded. North is up and East is
 to the left. The image covers $3^\prime\times2^\prime$.}
   \label{Fig.1}
\end{center}
\end{figure}

\bigskip

\begin{figure}[h]
\begin{center}
 \includegraphics[width=4.in]{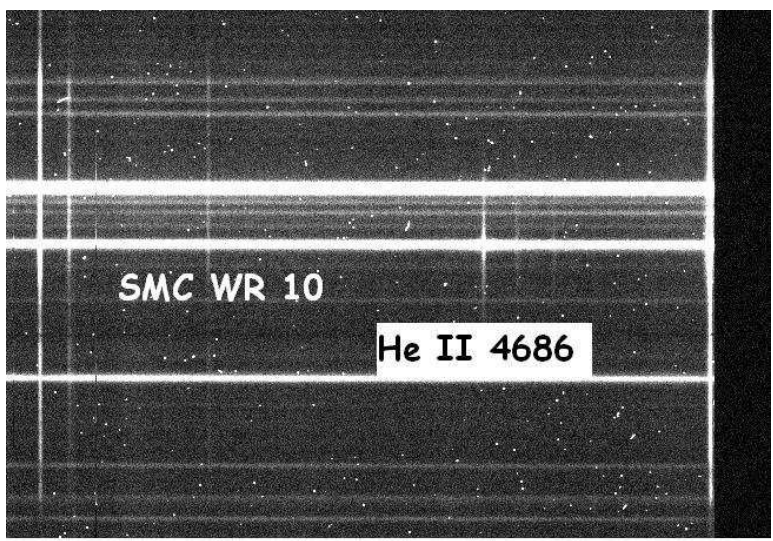} 
 \caption{\textit{ESO} \textit{NTT}/EMMI Long-slit spectrum of SMC\,WR10 and the 
 surrounding H\,{\sc ii} region NGC\,249. The strong vertical lines to the left and
 to the right are nebular H$\gamma$ and H$\beta$, respectively. The He\,{\sc ii}\,$\lambda$4686 
 emitting He\,{\sc iii} region extends $45^{\prime\prime}$ corresponding to some 10 pc 
 in the SMC.}
   \label{fig2}
\end{center}
\end{figure}

The particular importance of SMC\,WR10 for our understanding of the 
physics and evolution of WR stars comes from the fact that  this WN3ha
star is most likely single (\cite[Foellmi, Moffat \& Guerro
2003]{FMG03})  and therefore probably not a result of massive binary
evolution.  Moreover, the star  is faint ($V\sim 16.0$ mag), displays a
rather weak \mbox{($EW=25$ \AA)}  and narrow \mbox{($FWHM=30$ \AA)}
stellar He\,{\sc ii}\,$\lambda$4686 emission line (\cite[Crowther \&
Hadfield 2006]{CrHa06}), and it contains substantial amounts of hydrogen
in its atmosphere.

An important diagnostic for WRWs is the flux ratio, r$_{4686}$,  between
the narrow nebular component and the broad stellar He\,{\sc
ii}\,$\lambda$4686 emission component. Thus,  r$_{4686}$ directly
measures the fraction of He$^+$-ionizing radiation  that escapes the
Wolf-Rayet wind. For SMC\,WR10, one finds r$_{4686}\sim5$  which is
similar to the values derived for WRWs Brey\,2 and SMC\,WR7.     Fitting
this set of observations with the Potsdam WR model WNL grid  (cf.,
\cite[Hamann \& Gr\"afener 2004]{HaGr04}) one readily  derives the
following stellar parameters for SMC\,WR10:\\

$T_\star=100\pm10$ kK,\  $R_\star=2.4\pm0.3$ R$_{\odot}$,\  
$L=10^{5.7\pm0.2}$ L$_{\odot}$,\  $\dot{M}=2\ 10^{-6}$ M$_{\odot}/$yr\\

Note in particular the small derived mass-loss rate as compared to 
more typical Wolf-Rayet winds that are 10 to 100 times more powerful.
Interestingly, the \cite[Smith et al. (2002)]{Smetal02} CMFGEN
models suggest a significantly smaller temperature $T_\star=80$ kK
for the same He$^+$/H$^0$ Lyman continuum ratio. In other words, for the same
core temperature \mbox{CMFGEN} models predict roughly $10\times$ more He$^+$-ionizing
photons than the corresponding PoWR models. On my request, 
Wolf-Rainer Hamann kindly computed weak-wind PoWR models with exactly the 
same input parameters as the hot low metallicity (0.2 Z$_{\odot}$) stars
listed in \cite[Smith et al. (2002)]{Smetal02}. He found the same 
discrepancy and suggested that possibly different assumed gravities 
of the underlying cores might play a role here. 

As far as I am aware, no standard evolutionary track leads to objects
like SMC\,WR10. However, as mentioned earlier, for very rapid rotation
the evolution of massive stars is expected  to proceed  directly 
towards high effective temperatures (\cite[Yoon \& Langer
2005]{YL05})
into the region of the HR diagram where SMC\,WR10 is located. Quite
possibly, this is the first time that quasi-chemically homogeneous
evolution has been substantiated.    
 
\section{Highly excited 'blue' nebulae in the MCELS images}

Recently, Smith and the MCELS team (cf., \cite[Smith et al.
1999]{Smetal99})  published beautiful multi-colour images of the
Magellanic Clouds emphasizing nebular emission of  H$\alpha$ (red color
coded; $R$), [O\,{\sc iii}]\,$\lambda$5007 (blue; $B$),  and [S\,{\sc
ii}]\,$\lambda6716,31$ (green; $G$). These pictures are not only very
appealing (which explains why they have been shown in many talks during
the conference), but they also allow to readily discriminate between
various ionization/excitation mechanisms at work.

MCELS images easily allow to identify green and yellow ($y=R+G$)
shock-ionized SNR, often superimposed on normal reddish H\,{\sc ii}
regions, and they reflect the ionization structure of photoionized
nebulae that often turn "yellowish" and "brownish" ($=y+R$) towards the
edges  where recombination towards lower-ionization species occurs.

\begin{figure}[t!]
\begin{center}
 \includegraphics[width=4.0in]{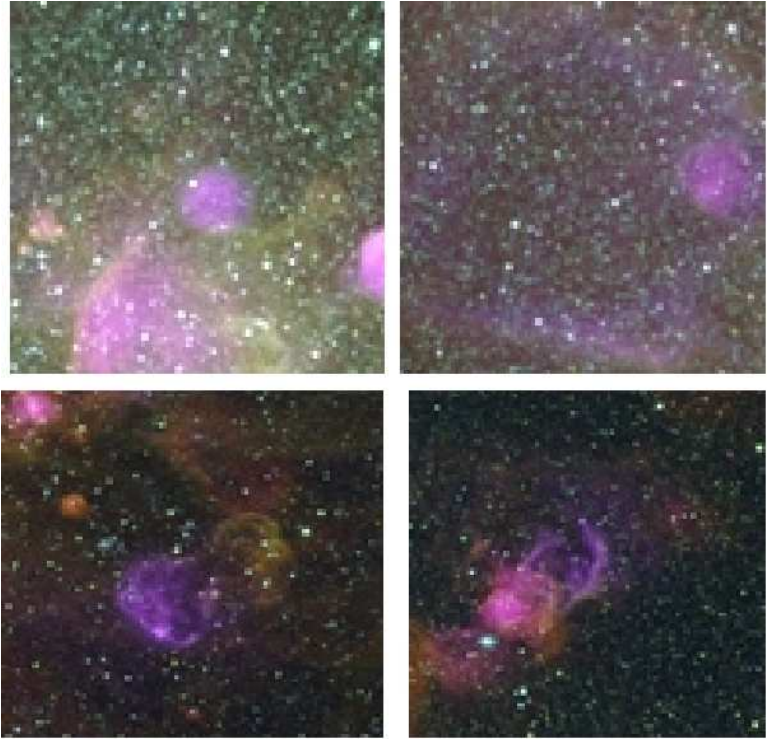} 
 \caption{Examples of strongly [O\,{\sc iii}]\,$\lambda$5007 emitting blue nebulae in the
 Magellanic Clouds. These cut-outs each measure about 
 $10^\prime\times10^\prime$ corresponding to 150 pc in the Clouds.
 One needs to look at the colour version of the images in order to 
 appreciate the unique "blueness" of the nebulae.
From upper left, clockwise: 
(a) small round nebula North of the H\,{\sc ii} region N\,19 (below) in the SMC, 
(b) previously unknown ring nebula around the faint WNE star SMC\,WR9, 
(c) a blue ring nebula in DEM\,66 (= N\,23A) in the LMC
(d) the high excitation blue nebula NGC\,1945 
in the LMC located next to low exitation nebula NGC\,1948 to the W (right)} 
\label{fig3}
\end{center}
\end{figure}

Even a casual inspection of the LMC image in particular readily reveals a class 
of very blue nebulae (hereafter BNe); i.e.\ nebulae that appear  to be
dominated by the [O\,{\sc iii}]\,$\lambda$5007 blue hue. I have
verified that the colour balance is such that nebulae appear 'very blue' when
the intensity ratio $\lambda$5007/H$\alpha$ $>1.5$; i.e., being of very high
excitation. I detect some 40 BNe in the LMC (not including the 30\,Dor region
which appears "burned-out") and about 10 BNe in the SMC, excluding NGC\,346  and
environment (cf.\ Fig.\ 3.)

What is the nature of the BNe? By employing the  Strasbourg {\sc Aladin
Sky Atlas} one quickly realizes that a large fraction  thereof
represents photoionized nebulae around known early WN stars (WNE), 
several of which  were not known before to excite ring nebulae around
them  (cf.\ the compilation by \cite{Doetal94}, and SMC WR update by
\cite{MaDu01}). Late WN (WNL) and WC types appear to create less highly
excited  (magenta-red coloured) nebulae more akin to normal O star
H\,{\sc ii} regions. The fact that such subtle  differences become so
strikingly apparent is indeed quite remarkable. It is also clear that
many WNE BNe are much more extended  than previously thought, and
several such bubbles have [O\,{\sc iii}]\,$\lambda$5007- 'skin' emission
due to photoionized/incomplete shocks  in the supersonically expanding
bubbles (see, e.g., \cite{Duf89}).

We then might ask the question: What is the nature of the BNe that are
not clearly associated with (known) WNEs? Remote possibilities include 
flash-ionized nebul\ae\ around young O-rich supernova remnants  (like in
the case of SNR\,0103$-$72.6), rare X-ray ionized nebulae (like around
LMC X-1 and Cal~83), or regions of  incomplete shocks at the periphery 
of more easily discernable low-excitation SNRs.  However, a more likely
possibility is that we see here the fingerprints of hidden  WN stars
that have but weak stellar emission lines, i.e., being even  fainer than
those of the recently identified class of weak-lined SMC WRs.  Such
stars could have easily escaped detection  by using slitless
spectroscopy or narrow-filter imaging techniques. Also, such objects
would have weak winds and therefore be prone to be WRW. A possible
example is the huge (150 pc diameter) newly detected  faint ring nebula
around SMC\,WR9 (a spectroscopic twin of WR\,10)  shown in Fig.\ 3. 

\section{A remark on the nature of He\,{\sc ii}\,$\lambda$4686 nebula N\,44C}    
 
Not unexpectedly, He\,{\sc iii} region N\,44C also appears as a
prominent BN in the MCELS image of the Large Cloud. However, its
ionizing  source has not been clearly identified, even though an
interpretation  in terms of a fossil X-ray ionized nebula cannot be
excluded.  Here I propose an alternative scenario, not least inspired by
the  variable radial velocity of the central O7 star (\cite[Pakull
1989]{Pa89b})  which clearly points to a binary nature. Indeed,
evolutionary calculations of massive binaries by  \cite[Wellstein,
Langer \& Braun (2001)]{Weetal01}  predict a sizable population of
massive O stars with {\it low-mass} (5--7 M$_\odot$), hot ($\sim10^5$ K)
Helium-star secondaries, after mass  transfer from the formerly more
massive component to its companion  has been completed.  The He stars
will be much fainter optically than the O star  primaries and will
therefore be quasi undetectable spectroscopically, readily explaining
the apparent absence of such systems.  However, the idea here is that
such a star will act as a Warmer, i.e., excite a He\,{\sc iii} region
that will in turn be detectable as a He\,{\sc ii}\,$\lambda$4686
emitting nebula provided that the interstellar  density is sufficiently
high. Conceivably, the central (binary) star in  N\,44C is such a
system.

\acknowledgements{
I thank the editors for their patience with the delivery of my manuscript.}


\begin{thebibliography}{}

\bibitem[Crowther \& Hadfield 2006]{CrHa06}
{Crowther, P.A., \& Hadfield, L.J.} 2006,
\textit{A\&A}, 449, 711

\bibitem[Dopita et al. (1994)]{Doetal94}
{Dopita, M.A., Bell, J.F., Chu, Y.-H., \& Lozinskaya, T.A.} 1994,
\textit{ApJS}, 93, 455

\bibitem[Dufour 1989]{Duf89}
{Dufour, R.J.} 1989, 
\textit{Rev. Mexicana AyA}, 18, 87

\bibitem[Foellmi, Moffat \& Guerro (2003)]{FMG03}
{Foellmi, C., Moffat, A.F.J., \& Guerro, M.A.} 2003,
\textit{MNRAS}, 338, 360 

\bibitem[Hamann \& Gr\"afener (2004)]{HaGr04}
{Hamann, W.-R., \& Gr\"afener, G.} 2004,
\textit{A\&A}, 427, 697 --- http://www.astro.physik.uni-potsdam.de/$\sim$wrh/PoWR/powrgrid1.html

\bibitem[Massey \& Duffy (2001)]{MaDu01}
{Massey, P., \& Duffy, A.S.} 2001, 
\textit{ApJ}, 550, 713

\bibitem[Niemela et al. (1991)]{Nie91}
{Niemela, V.S., Heathcote, S.A., \& Weiller, W.C.} 1991, 
in K.A. van der Hucht \& B. Hidayat (eds.), 
\textit{Wolf-Rayet stars and interrelations with other massive stars in galaxies}, IAU Conf.Proc. 143 (Dordrecht: Kluver), p.\,425

\bibitem[Pakull\& Motch 1989a]{PaMo89}
{Pakull, M.W. \& Motch, C.} 1989a,
\textit{Nature} 337, 337

\bibitem[Pakull \& Motch (1989b)]{Pa89a}
{Pakull, M.W., \& Motch, C.} 1989b, in E.J.A. Meurs \& R.A.E. Fosbury (eds.), 
\textit{ESO workshop on extranuclear activity in galaxies} (Garching: ESO), p.\,285 

\bibitem[Pakull (1989)]{Pa89b}
{Pakull, M.W.} 1989, in K.S. de Boer, F. Spite, \& G. Stasi\'nska (eds.), 
\textit{Recent developments in Magellanic Cloud research} (Meudon: Observatoire de Paris), p.\,183

\bibitem[Pakull (1991)]{Pa91}
{Pakull, M.W.} 1991, in K.A. van der Hucht \& B. Hidayat (eds.), 
 \textit{Wolf-Rayet stars and interrelations with other massive stars in galaxies}, IAU Conf.Proc. 143 (Dordrecht: Kluver), p.\,391

\bibitem[Pakull \& Bianchi (1991)]{Pabi91}
{Pakull, M.W., \& Bianchi, L.} 1991, in K.A. van der Hucht \& B. Hidayat (eds.), 
 \textit{Wolf-Rayet stars and interrelations with other massive stars in galaxies}, IAU Conf.Proc. 143 (Dordrecht: Kluver), p.\,260

\bibitem[Pakull \& Angebault (1986)]{PA86}
{Pakull, M.W, \& Angebault, L.P.} 1986,
\textit{Nature}, 322, 511 

\bibitem[Schmutz, Leitherer \& Gruenwald (1992)]{SLG92}
{Schmutz, W., Leitherer, C., \& Gruenwald, R.} 1992,
\textit{PASP}, 104, 1164 

\bibitem[Smith et al. (1999)]{Smetal99}
{Smith, R.C., \& the MCELS Team} 1999, in Y.-H. Chu, N. Suntzeff, J. Hesser, \& D. Bohlender (eds.), \textit{New views of the Magellanic Clouds}, IAU Conf.Proc. 190, p.\,28 --- http://www.ctio.noao.edu/$\sim$mcels/

\bibitem[Smith et al. (2002)]{Smetal02}
{Smith, L.J., Norris, R.P.F., \& Crowther, P.A.} 2002,
\textit{MNRAS}, 337, 1309 

\bibitem[Terlevich \& Melnick (1985)]{TM85}
{Terlevich, R., \& Melnick, J.} 1985,
\textit{MNRAS}, 213, 841 


\bibitem[Vink \& deKoter (2005)]{ViKo05}
{Vink, J.S., \& de Koter, A.} 2005,
\textit{A\&A}, 442, 587

\bibitem[Wellstein, Langer \& Braun (2001)]{Weetal01}
{Wellstein, S., Langer, N., \& Braun, H.} 2001,
\textit{A\&A}, 369, 939 

\bibitem[Yoon \& Langer (2005) ]{YL05}
{Yoon, S.-C., \& Langer, N.} 2005,
\textit{A\&A}, 443, 643 

\end{thebibliography}
\end{document}